\documentclass[prb,
twocolumn, superscriptaddress,showpacs,amsmath,amssymb]{revtex4}
\usepackage{amsfonts}
\usepackage{bm}
\usepackage{verbatim}
\usepackage{xcolor}

\usepackage{graphicx}

\newcommand{\im}{i}
\newcommand{\abs}[1]{\vert #1 \vert}
\newcommand{\unitvec}[1]{\hat{\mathbf{#1}}}

\begin{document}

\title{Dimensional crossover of effective orbital dynamics in polar distorted $^3$He-A: Transitions to anti-spacetime}

\author{J.~Nissinen}
\affiliation{Low Temperature Laboratory, Aalto University,  P.O. Box 15100, FI-00076 Aalto, Finland}

\author{G.E.~Volovik}
\affiliation{Low Temperature Laboratory, Aalto University,  P.O. Box 15100, FI-00076 Aalto, Finland}
\affiliation{Landau Institute for Theoretical Physics, acad. Semyonov av., 1a, 142432,
Chernogolovka, Russia}

\date{\today}

\begin{abstract}
Topologically protected superfluid phases of $^3$He allow one to simulate many important aspects of relativistic quantum field theories and quantum gravity in condensed matter.
Here we discuss a topological Lifshitz transition of the effective quantum vacuum in which the determinant of the tetrad field changes sign through a crossing to a vacuum state with a degenerate fermionic metric. Such a transition is realized in polar distorted superfluid $^3$He-A in terms of the effective tetrad fields emerging in the vicinity of the superfluid gap nodes: the tetrads of the Weyl points in the chiral A-phase of $^3$He and the degenerate tetrad in the vicinity of a Dirac nodal line in the polar phase of $^3$He. The continuous phase transition from the $A$-phase to the polar phase, i.e. in the transition from the Weyl nodes to the Dirac nodal line and back, allows one to follow the behavior of the fermionic and bosonic effective actions when the sign of the tetrad determinant  changes, and the effective chiral space-time transforms to anti-chiral "anti-spacetime".
This condensed matter realization demonstrates that while the original fermionic action is analytic across the transition, the effective action for the orbital degrees of freedom (pseudo-EM) fields and gravity have non-analytic behavior. In particular, the action for the pseudo-EM field in the vacuum with Weyl fermions (A-phase) contains the modulus of the tetrad determinant. In the vacuum with the degenerate metric (polar phase) the nodal line is effectively a family of $2+1$d Dirac fermion patches, which leads to a non-analytic $(B^2-E^2)^{3/4}$ QED action in the vicinity of the Dirac line. 
\end{abstract}
\pacs{
}

\maketitle

\section{Introduction}

Recent research on topological phases of matter has focused on gapless fermionic quasiparticles topologically protected in the bulk of the system. Examples include Dirac and Weyl fermions and their possible higher spin generalizations in non-relativistic condensed matter systems not constrained by Lorentz symmetry \cite{BradlynEtAl16,Winkler16,Fulga2017}. Such quasiparticles feature prominently in e.g.  semi-metals,\cite{Herring1937,Abrikosov1971,Abrikosov1972} topological superfluids and superconductors where they are protected by topology and/or various discrete symmetries. What sets such systems apart from other gapless (massless) fermionic systems in condensed matter and high-energy physics is the emergent quasirelativistic form of the low-energy dispersion.\cite{FrogNielBook,Horava2005,Volovik2003} The symmetries of the low-energy dispersion, i.e. the effective fermionic metric, are set by the low-energy ground state properties and background fields and thus are not typically constrained by Lorentz symmetry. Among the most striking features of such topologically protected gapless systems is the presence of quantum anomalies similar to relativistic field theories \cite{NielsenNinomiya83, Volovik90, Bevan1997, KlinkhamerVolovik05, RyuMooreLudwig12, ZyuzinBurkov12, BurkovKim16, GoothEtAl17, Burkov17} and surface states with dispersionless bands, e.g. Fermi arcs \cite{WanEtAl2011}, drumhead states and flat bands \cite{BurkovHookBalents11, HeikkilaEtAl11, Heikkila2015}. 
 
As in any fermionic theory, the coupling of the low-energy fermions to geometry and gauge fields is through effective tetrad fields (and the spin connection). For example, the determinant of the tetrad $\det e$ may have different signs throughout the system (as dictated e.g. by global topology), but the metric and dispersion are not sensitive to this sign. On the other hand, the presence of topological terms in the effective action after the low-energy fermions have been integrated out can make the theory anomalous under such transformations. One is therefore interested in the global spacetime symmetries $P,T$ and other discrete symmetries of the system that can be anomalous in the low-energy quantum theory. 

Here we address the question what happens
if $\det e$ changes sign and the right-handed fermions continuously transform to the left-handed fermions and {\it vice versa} (i.e. not necessarily through a symmetry operation) \emph{without} considering the possible anomalous or topological terms in the effective action. The relativistic analogue of such a transformation was termed ``anti-spacetime" in Ref. \onlinecite{Rovelli2012b} and has been considered e.g. in Refs. \onlinecite{Rovelli2012a,Diakonov2011,Diakonov2012}. As the tetrad and metric become singular as $\det e$ crosses zero, this connects to the problem in quantum gravity over which type of singular metrics should be integrated over in the gravitational path integral \cite{DAuriaRegge82}. Ref. \onlinecite{Rovelli2012b} suggested three alternatives to this problem: (i) Anti-spacetime does not
exist, and ${\rm det} \,e>0$ should constrain the gravity path integral;
(ii) Anti-spacetime exists, but the action depends on $|{\rm det} \,e|$, rather than on  ${\rm det} \,e$. (iii) Anti-spacetimes exists and contribute nontrivially to quantum gravity.

We consider the problem of degenerate tetrads for Weyl and Dirac fermions in superfluid $^3$He in the A- and polar phases, respectively. Related to the singular tetrads, we study the resulting low-energy effective actions for the orbital degrees of freedom of the superfluid. Formally these couple to the (neutral) $^3$He fermions in the form of effective 3+1d and 2+1d QED. Namely we will focus on the effective Euler-Heisenberg actions \cite{Dunne05, Redlich1984, Volovik92} for the orbital dynamics. In particular, for the Weyl fermions in $^3$He-A, we can make a transformation to effective anti-spacetime with $\det e<0$ through the (topological Lifshitz) phase transition to the polar phase of $^3$He, or more trivially through a continuous $SO(3)$ rotation of the orbital angular momentum anisotropy axis in the laboratory frame (see Fig. \ref{transition}). In the polar phase, the full 3+1d tetrad is degenerate, i.e. ${\rm det} \,e=0$, and the low-energy fermionic spectrum is characterized by  a nodal line of 2+1d Dirac fermions. We then consider the singularities in the effective action which occur due to the presence of the nodal line. In this case the $3+1$d spacetime for fermions near the Weyl point transforms to a family of $2+1$d  effective low-energy spacetimes, which leads to a non-analytic effective action for the corresponding pseudo-EM orbital fields emerging in the vicinity of the nodal line. 

The transformation from chiral $^3$He-A spacetime to chiral anti-spacetime through the polar phase in $^3$He demonstrates that the emerging bosonic low-energy effective action  does not depend on the sign of  $\det e$, instead it contains  $\sqrt{-g}=\vert \det e \vert$ and is therefore non-analytic in terms of the tetrad field. The transition with a degenerate tetrad also demonstrates that the original action for fermions is analytic, but it is not fully diffeomorphism invariant. It is invariant under the  volume-preserving (i.e. $\delta g = 0$) coordinate transformations. By rescaling, it can be made  diffeomorphism invariant, but after that it becomes non-analytic. The emergent bosonic effective action, obtained by integration over fermions, acquires the same properties: it is invariant under coordinate transformations but is non-analytic in terms of the tetrad. Depending on the details, it is possible that topological terms emerge, they are however, anomalous under discrete (or gauge) symmetry transformations.

Finally, the topological Lifshitz  transition associated with the change of sign of  $\det e$ has been also considered in for the so-called type-III and IV interacting (or driven) 
Weyl fermions \cite{NissinenVolovik2017b} with tilted Weyl cones in frequency space. Here we discuss this transition for the interacting (or driven) type-II Weyl fermions with a non-zero tilt of the conical dispersion in frequency \cite{HuhtalaVolovik2002,VolovikZubkov2014,Soluyanov2015}. In this case the intermediate state at the Lifshitz transition contains a degeneracy surface of the bands touching at the Weyl point instead of the nodal line in the spectrum \cite{mcclure57,Mikitik2006,Mikitik2008,Heikkila2015,Heikkila2016}.

\section{Polar distorted $^3$He-A}

According to von Neumann and Wigner,\cite{Neumann1929}, in 3+1 dimensions two bands may touch each other at a point (Weyl or Dirac point of co-dimension 3) or have a band degeneracy along a line (a Dirac nodal line of co-dimension 2). The topological origin of the band touching is discussed e.g. in Ref.\cite{Novikov1981}. Weyl points are topologically protected by an integer valued topological invariant $N_3$ in momentum space, see e.g. \cite{Volovik2003} and a zero-sum-rule $\sum_a N_{3,a} = 0$ of total chirality applies in the bulk of any system \cite{NielsenNinomiya83}. The existence of Weyl points requires broken $P$ or $T$ symmetry, while the combination $PT$ remains a symmetry. 
Violation of $PT$  symmetry leads to inflation of the Weyl points and formation of the Fermi pockets.\cite{Volovik1989,Volovik2003,Timm2017}  Examples of Weyl points are found in topological superfluids, superconductors and semi-metals: two spin-degenerate $N_3= \pm2$ Weyl points are found both in $^3$He-A and in the distorted $^3$He-A that break time-reversal symmetry. Similarly, the Dirac nodal line is found in the Bogoliubov spectrum of the recently experimentally discovered polar phase of $^3$He\cite{Dmit1,Autti2016}, where the nodal line is protected by topology together with time-reversal symmetry $T$. Throughout the paper, we shall focus on these topological spin-triplet $p$-wave superfluids with gapless Dirac and Weyl  fermions.

\subsection{Weyl fermions in polar distorted $^3$He-A}
The spin-triplet $p$-wave order parameter $\Delta_{\alpha\beta}(\mathbf{k}) = (\im \sigma^2 \sigma^{\mu})_{\alpha\beta}k^iA_{\mu i}$ of polar distorted superfluid $^3$He-A is given by \cite{Volovik92}
\begin{align}
A_{\mu i} =   \hat{d}_{\mu}(m_i + i n_i), \label{eq:OPHeA}
\end{align}
where $\unitvec{d}$ is the spin-anisotropy axis and the orbital structure is parametrized by two real and orthogonal vectors, $\mathbf{m} \cdot \mathbf{n} = 0$.
It is convienient to introduce the third unit vector $ \hat{\bf l}$:
\begin{equation}
 \hat{\bf l}= \frac{{\bf m}  \times{\bf n} } {| {\bf m} \times{\bf n}|}    \,.
\label{l}
\end{equation}
If $|\mathbf{m}|  = |\mathbf{n}| = \Delta_0/p_F$, Eq. \eqref{eq:OPHeA} describes $^3$He-A with gap $\Delta_0$. The chiral A-phase Cooper pairs have orbital angular momentum along $\hat{\bf l}$ and therefore the phase has uniaxial orbital anisotropy along this unit vector. If $\mathbf{n}=0$ (or $\mathbf{m}=0$), one has the polar phase, which is time reversal invariant. Otherwise, Eq. \eqref{eq:OPHeA} describes the polar distorted $A$-phase \cite{Dmit1}.

\begin{figure*}[htb!]
\includegraphics[width=1.0\linewidth]{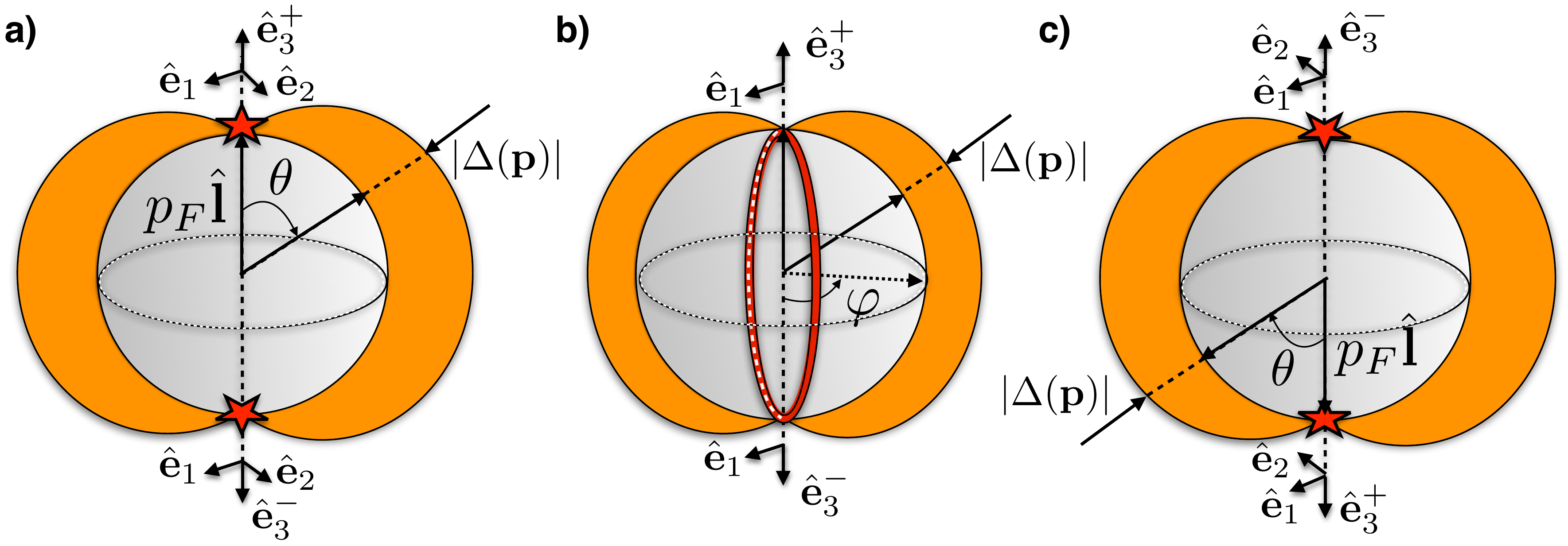}
\caption{Schematic illustration of the Lifshitz transition through the polar phase in polar distorted $^3$He-A, where right-handed fermions transform to left-handed and {\it vice versa}. a)  $^3$He-A has two (spin-degenerate) Weyl points of opposite chirality (red stars) at $\mathbf{W}_{\pm} = \pm p_F\mathbf{l}$ with triads $\mathbf{e}_a^{\pm}$. b) As the triad $\mathbf{e}_{2}$ is varied adiabatically through zero, the polar phase with a nodal line (red) emerges from the poles when $\mathbf{e}_2 =0$. c) For the adiabatic evolution $\unitvec{e}_2 \to -\unitvec{e}_2$ or a $SO(3)$-rotation of the whole orbital part $\unitvec{l} \to -\unitvec{l}$ with respect to the laboratory frame, the chirality of the Weyl points interchanges as well. 
}
 \label{transition}
\end{figure*}

The model, weakly-coupled Bogoliubov-de Gennes Hamiltonian corresponding to Eq. \eqref{eq:OPHeA} for the fermionic quasiparticles is given by
\begin{align}
H(\mathbf{p}) = \frac{p^2 - p_F^2}{2m}\tau^3 + \mathbf{m}  \cdot \mathbf{p} \,\tau^1 + \mathbf{n}  \cdot \mathbf{p} \,\tau^2  
\, 
\label{HamiltonianA}
\end{align}
where $\tau^a$ are Pauli matrices in Bogoliubov-Nambu space and the spin quantum number along $\unitvec{d}$ is degenerate and ignored. 
In the A-phase and in the  polar distorted A-phase there are two Weyl points  located at momenta $\mathbf{W}_{\pm} = \pm p_F\unitvec{l}$ in Fig.\ref{transition}.
The low-energy Hamiltonians for quasiparticles in the vicinity of these Weyl points,  ${\bf W}_+$  and ${\bf W}_-$,  have the form:
\begin{align}
H^{\pm} ({\bf p})=  ({\bf p} - {\bf W}_{\pm})\cdot ( {\bf e}_1\tau^1 + {\bf e}_2\tau^2  
+ {\bf e}_3^{\pm} \tau^3) \,,
 \label{HamiltonianWeyl}
\end{align}
where
\begin{equation}
 {\bf e}_1 = {\bf m} \,, \,   {\bf e}_2 = {\bf n}, \quad  {\bf e}_3^{\pm}=\pm v_F \unitvec{l} \,,
\label{triads}
\end{equation}
and  $v_F=p_F/m$.
The BdG Hamiltonians Eq. \eqref{HamiltonianWeyl} describe fermions with an effective tetrad field $e^k_{a}$, effective gauge field $\mathbf{A}_{\rm eff}$, and charges $q_{\pm} = \pm 1$:
\begin{align}
H = \mathbf{e}_{a}\tau^a\cdot(\mathbf{p}-q_{\pm}\mathbf{A}), \quad
e^{k}_a = (\mathbf{e}_a)^k, \quad \bold{A}_{\rm eff} = p_F \unitvec{l}
\,.
 \label{eq:HamiltonianWeyl}
\end{align}
Note that the two tetrad fields $e^k_{a}$ in Eq.(\ref{triads}) have the antiparallel components $\mathbf{e}^{\pm}_3$, and thus  the tetrad determinants for these two Weyl fermions have opposite signs:
\begin{equation}
e_{\pm} =\pm |\mathbf{m}|   |\mathbf{n}| v_F.
\label{e}
\end{equation}
This means that these fermions have opposite chiralities, while they have the same effective metric $g^{ik}= e_1^i e_1^k +  e_2^i e_2^k+  e_3^i e_3^k$  with determinant:
\begin{equation}
 \sqrt{-g}=|e_+|^{-1}=  |e_-|^{-1}= \frac{1}{  |{\bf e}_1||{\bf e}_2| |{\bf e}_3| }  \,,
\label{g}
\end{equation}
The chirality of fermions is determined by  
the orientation of the unit vector $\hat {\bf l}$: 
The Weyl fermions at ${\bf W}_{+}=+p_F\hat {\bf l}$ are always right-handed and the topological charge $N_3=+1$ describes a hedgehog with (pseudo)spins pointing outward.  The Weyl fermions at ${\bf W}_-=-p_F\hat {\bf l}$ are always left-handed with the topological charge $N_3=-1$ \cite{Volovik2003}. 

\subsection{Change of sign of tetrad determinant across the transition through the polar phase}
Let us discuss the Lifshitz transition where $\mathbf{n}=\mathbf{e}_2= b\hat{\bf y}$ and the parameter $b$ crosses zero with $1 \geq b\geq -1$ \cite{Dmit1}, see Fig. \ref{transition}. At $b=0$ the determinant of the tetrad changes sign and the right-handed fermions transform to the left-handed ones\cite{Diakonov2011,Rovelli2012a,Rovelli2012b}. At this transition, the polar phase where the two fermion bands are degenerate along a line, is crossed. From the point of view of the low-energy fermions and effective gravity, the 3+1d tetrad is degenerate with $\det e = 0$ in the polar phase, corresponding to a nodal line of $2+1$d spacetimes in the transverse directions in momentum space (we ignore here the tiny symmetry-breaking spin-orbit interaction which destroys the Dirac nodal line\cite{Zubkov2017}).

Across this transition the $\hat {\bf l}$-vector changes its orientation to its opposite, $\mathbf{W}_{\pm} \to \mathbf{W}_{\mp}$, and the chiralities at  $\mathbf{p} = {\bf W}_+$ and  $\mathbf{p} = {\bf W}_-$ points in momentum space change sign, compare  Fig.\ref{transition}(a) and  Fig.\ref{transition}(c).
In Fig. \ref{transition} (a) the point ${\bf W}_+ = p_F\hat {\bf l}$ is the right-handed Weyl point,
while after transition in Fig.\ref{transition}(c) it is expressed in terms of  $\hat {\bf l}$ as ${\bf W}_- = -p_F\hat {\bf l}$, and thus becomes the left-handed Weyl point. At  ${\bf e}_2=0$, i.e. in the polar phase, the spatial part of the $3+1$ tetrad $e^{\mu}_{a}$ becomes degenerate, which signals appearance of the 2+1d nodal line. According to the analysis Ho\v{r}ava \cite{Horava2005} and the topological classification of Refs. \cite{ZhaoWang13, ZhaoSchnyderWang16}, one expects that relativistic invariance emerges in the low-energy transverse directions. The system effectively transforms to a one-parameter family of  $2+1$ QFTs, with dyads ${\bf e}_1$ and ${\bf e}_3$ along the transverse directions that depend on the position along the nodal line in Fig.\ref{transition}(b).

\subsection{Effective 3+1d QED}

Let us first consider the effective electrodynamics of the orbital degrees of freedom for the case of the $3+1$d polar distorted A-phase, when approaching the polar phase with $\mathbf{e}_2=0$. The position of the Weyl point plays the role of pseudo gauge field, ${\bf A}_{\rm eff}=p_F\hat{\bf l}$, with the corresponding pseudo magnetic field $F_{ik}=\nabla_i A_k - \nabla_k A_i=p_F\epsilon_{ikm}(\nabla \times\hat{\bf l})_m$. For the static case, $\partial_t \mathbf{A}_{\rm eff} = 0$, the effective free energy obtained after integration over the massless Weyl fermions has logarithmic contribution, see e.g. \cite{Volovik92, Volovik2003}:
\begin{equation}
\Delta F_{\rm grad}[B] = \frac{\sqrt{-g} }{48\pi^2} B^2\ln \left( \frac{E_{\rm UV}^2}{B} \right) + \cdots\,,
\label{FFS1}
\end{equation}
where $B = \sqrt{B^2}$ is the quasirelativistic contraction of the pseudo magnetic field,
\begin{equation}
B^2 = g^{im}g^{kn} F_{ik}F_{mn}\,,
\label{FF}
\end{equation}
and $E_{\rm UV}$ is a cutoff scale of the order of the superfluid gap $\Delta_0$ and the ellipses represent subleading contributions compared to $\log \times B^2$ as $B\to 0$. In superfluid $^3$He-A, one has $|{\bf e}_1|=|{\bf e}_2|=\Delta_0/p_F \ll v_F =|{\bf e}_3|$, and neglecting the terms of the relative magnitude $\Delta_0/v_Fp_F\ll 1$ one obtains
\begin{align}
\Delta F_{\rm grad} 
= \frac{p_F^2v_F \ln \left( \frac{E_{\rm UV}^2}{B} \right)}{48\pi^2 |{\bf e}_1||{\bf e}_2|} 
( ({\bf e}_1\cdot(\hat{\bf l}\cdot\nabla)\hat{\bf l})^2+ ({\bf e}_2\cdot(\hat{\bf l}\cdot\nabla)\hat{\bf l})^2)
.
\label{FFS2}
\end{align}

\subsection{Singularity in the effective 3+1d QED near the degeneracy of the metric. }

It is important that both Weyl points of opposite chirality give the same contribution to the effective action. This means that the action depends only on the modulus of the tetrad determinant $|e_{\pm}|$ and reflects the fact that the polar phase with $\det e=0$ is singular from the perspective of the effective $3+1$d spacetime and QED: the metric becomes degenerate and the effective action diverges. Similarly, when  ${\bf e}_2$ crosses zero, the sign of the tetrad determinant changes and the left-handed Weyl fermions transform to right-handed ones in the polar distorted $^3$He-A phase following after the transition.

The singularity in the effective action can be seen from Eq.(\ref{FFS2}): When the polar state with a degenerate tetrad approaches, i.e. when, say ${\bf e}_2\rightarrow 0$, one has:
\begin{align}
F_{\rm grad} \rightarrow \frac{p_F^2v_F \ln \left(\frac{E_{\rm UV}^2}{B}\right) }{48\pi^2}  \frac{1}{ |{\bf e}_1||{\bf e}_2|} 
 ({\bf e}_1\cdot(\hat{\bf l}\cdot\nabla)\hat{\bf l})^2
\\
= \frac{p_F^2v_F \ln \left(\frac{E_{\rm UV}^2}{B}\right) }{48\pi^2}  \frac{ |{\bf e}_1|}{|{\bf e}_2|} 
 (\hat{\bf l}\cdot(\hat{\bf l}\cdot\nabla)\hat{\bf m})^2\,.
\label{FFSlimit}
\end{align}
Here we explicitly introduced the unit vector $\unitvec{m}$ along ${\bf e}_1$, i.e. ${\bf e}_1 = \vert \mathbf{e}_1 \vert \unitvec{e}_1= \frac{\Delta_0}{p_F}\hat{\bf m}$ and we took into account that $\hat{\bf l}\cdot\hat{\bf m}=0$. The divergence of the quadratic term in Eq.(\ref{FFSlimit}) suggests that at the degenerate point of the tetrad, i.e. in the polar phase, where ${\bf e}_2=0$, the low-energy effective action with lower power of the gradients emerges. Moreover, in the polar phase one has a nodal line of zeroes and near the nodal line, the low-energy fermions are effectively 2+1d Dirac fermions rather than 3+1d Weyl fermions near the point nodes. For $2+1$d QED with massless fermions, the leading singularity of the effective low-energy action  comes with a $B^{3/2}$ term instead of the $B^2\ln(E_{\rm UV}^2/B)$, see e.g.  Ref. \onlinecite{Redlich1984}. The analogous effective action for the orbital dynamics of the polar phase is discussed in the next section.

Note that the original fermionic action and the Hamiltonian in Eq.(\ref{HamiltonianA}) 
 do not experience any singularity at the transition to the polar phase when ${\bf e}_2={\bf m}=0$. However, $H(\mathbf{p})$ has axial anisotropy along $\unitvec{l}$ and the theory is not (classically) diffeomorphism nor even Lorentz invariant. 
  The low-energy fermions in Eq. \eqref{eq:HamiltonianWeyl}, which acquire Lorentz invariance,  also experiences no singularity at the transition through ${\bf e}_2=0$. However,
Eq. \eqref{eq:HamiltonianWeyl} still lacks the full diffeomorphism invariance, obeying only the invariance under unimoduar coordinate transformations. 
 After the introduction of new tetrads  $\tilde{\bf e}_1=|e|^{-1/3}{\bf e}_1$, $\tilde{\bf e}_2=|e|^{-1/3}{\bf e}_2$,
 $\tilde{\bf e}_3=|e|^{-1/3}{\bf e}_3$ and $e^0_0=|e|^{-1/3}$ with $|\tilde e| = |e|^{1/3}$, the action with the Hamiltonian 
\begin{align}
\tilde{H}^{A}(\mathbf{p}) = |\tilde e| \,\tilde{\mathbf{e}}_a \tau^a \cdot (\mathbf{p}- q_A \mathbf{A}_{\rm eff})\,,
\end{align}
 becomes invariant under the full coordinate transformations.
However, it contains a fractional power of the original tetrad determinant.  As a result the action becomes non-analytic, in contrast to the original fermionic action. This explains the origin of the non-analyticity in $e$ in the bosonic action: it can be analytic only in theories of unimodular gravity\cite{vanderBij1982,Smolin2009,Klinkhame2017,KlinkhamerVolovik2016}. 
 
\section{Polar phase, degenerate metric, non-analytic 2+1-dimensional QED}

\subsection{Dirac fermions and $2+1$d QED in the polar phase}

\begin{figure}
\includegraphics[width=0.95\linewidth]{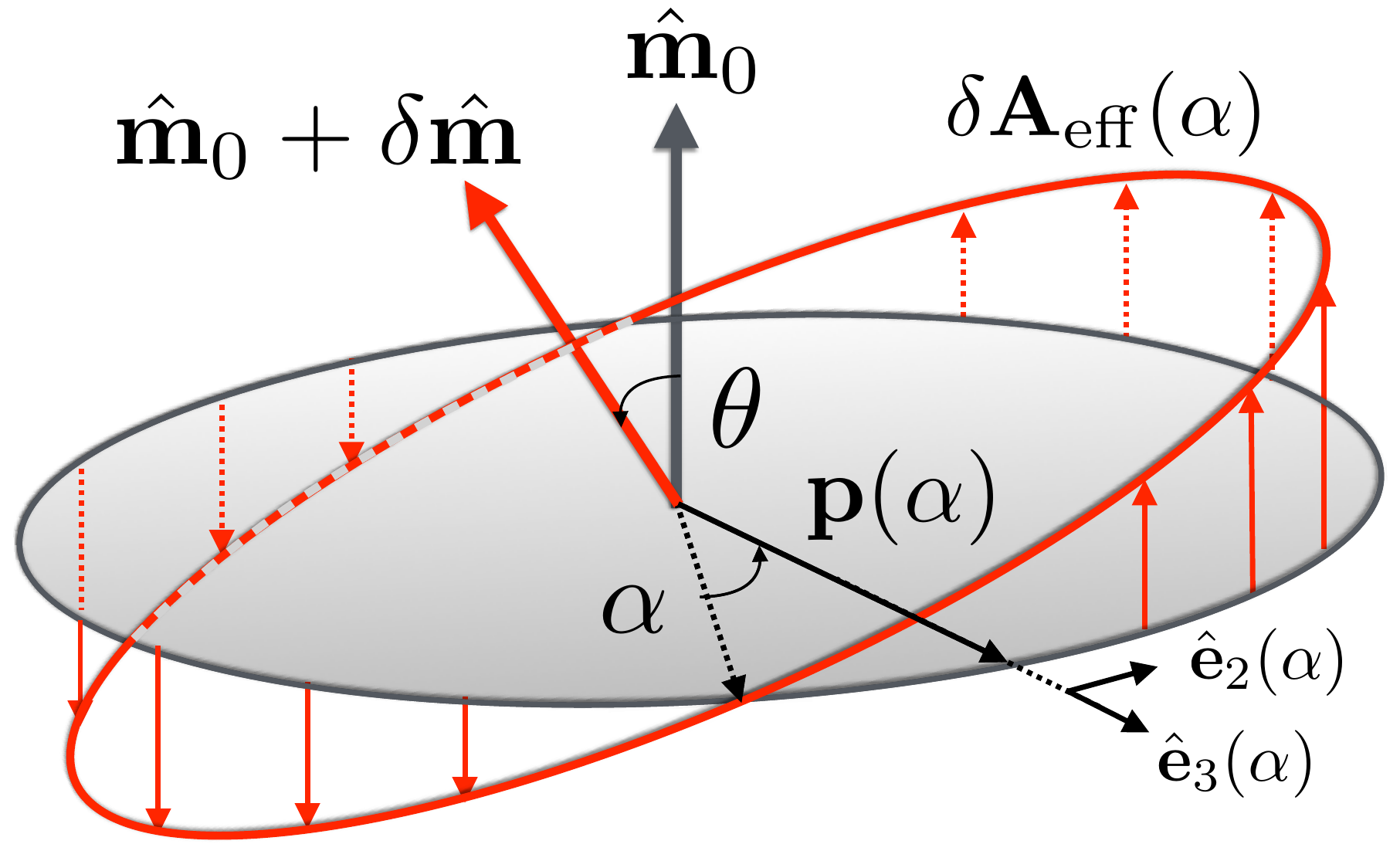}
\caption{Dependence of the effective vector potential of $2+1$ Dirac fermions on their position on the nodal line in Eq.(\ref{A}). This field is caused by the texture of the polar phase order parameter $\hat{\bf m}({\bf r})=\hat{\bf m}_0 + \delta\hat{\bf m}({\bf r})$. This vector is perpendicular to the nodal line, and thus the deformation $\hat{\bf m}({\bf r})$   tilts the plane of the nodal line. Such deformation corresponds to the Goldstone mode propagating in the polar phase. In terms of the effective 2+1d QED, this is the photon mode.}
\label{fig:deformation}
\end{figure}

When ${\bf n}=0$ in Eq.\eqref{eq:OPHeA}, one has the polar phase with ${\bf e}_2=0$ and the BdG Hamiltonian with a nodal line,
\begin{align}
H({\bf p})= v_F(p - p_F)\tau^3  +\mathbf p \cdot \mathbf{e}_1\tau^1 
\label{HamiltonianPolar}
\end{align}
where ${\bf e}_1= c_{\perp}\hat{\bf m}$ with $c_{\perp} = \frac{\Delta_0}{p_F}$. For $\mathbf{e}_2 = 0$, the two  Weyl points transform to  the nodal line at 
${\bf p} \cdot \unitvec{m}=0$, $p^2=p_F^2$. Let us parametrize the nodal line and the normal vector as a function of angle $\alpha$, see Fig. \ref{transition}(b) and Fig. \ref{fig:deformation}:
\begin{equation}
{\bf p}(\alpha)= p_F(\hat{\bf l}\cos\alpha +  \hat{\bf l} \times \hat{\bf m}\sin\alpha) =m^*  {\bf e}_{3}(\alpha)\,,
\label{line}
\end{equation}
here $\unitvec{l}$ is a unit vector perpendicular to $\hat{\bf m}$, and $m^*=p_F/v_F$. The unit vector $\unitvec{e}_2(\alpha) = (\unitvec{l}\times \unitvec{m}) \cos \alpha  - \unitvec{l} \sin \alpha$ is along the tangent of the nodal line and perpendicular to $\unitvec{p}(\alpha)$. 

For each $\alpha$, one has a patch of 2+1d massless Dirac fermions, with the low-energy Hamiltonian
\begin{equation}
H({\bf p};\alpha)=  ({\bf p}- {\bf p}(\alpha))\cdot ({\bf e}_1\tau^1 +{\bf e}_{3}(\alpha)\tau^3)  \,.
\label{HamiltonianLine}
\end{equation}
Note that $\mathbf{p}(\alpha+\pi) = -\mathbf{p}(\alpha)$, and the point antipodal to $\alpha$ has a Dirac fermion with opposite chirality. We again identify $\mathbf{A}_{\rm eff}(\alpha) = \mathbf{p}(\alpha)$ as the effective pseudo EM field coupled to the low-energy fermions. To study the fluctuations of the nodal line, we proceed as in the previous section by constructing the effective action for $\mathbf{A}_{\rm eff}(\alpha)$.
This corresponds to effective orbital gradient energy of the orbital order parameter $\hat{\bf m}$ of the polar phase induced by the quantum fluctuations of the superfluid vacuum. 

Let us parametrize the fluctuations of the nodal line induced by a texture of the superfluid order parameter $\unitvec{m}(\mathbf{r})$ as follows. Rotations of the polar axis $\unitvec{m}$ are Goldstone modes of the superfluid vacuum (in experiments, $\unitvec{m}$ is fixed by the uniaxial anisotropy of the aerogel sample \cite{Dmit1}). Set $\unitvec{m}_0 = \unitvec{z}$ and a perpendicular unit vector $\unitvec{l}_0=\unitvec{x}$ in a homogenous equilibrium state. Without loss of generality $\unitvec{m}(\mathbf{r}), \unitvec{l}(\mathbf{r})$ can be chosen as (see Fig. \ref{fig:deformation}):
\begin{align}
\hat{\bf m}({\bf r}) &\equiv \unitvec{m}_0+\delta \unitvec{m} = \hat{\bf z}\cos\theta({\bf r}) + \hat{\bf x}\sin\theta({\bf r}) \label{deformation} \\
&\approx \hat{\bf z}+ \hat{\bf x} \theta(\mathbf{r}), \nonumber \\
\hat{\bf l}({\bf r}) &\equiv \unitvec{l}_0+\delta \unitvec{l} = \hat{\bf x}\cos\theta({\bf r}) -\hat{\bf z}\sin\theta({\bf r}) \\
&\approx \hat{\bf x} - \hat{\bf z} \theta(\mathbf{r}), \nonumber 
\end{align}
where the slowly varying rotation parameter $\theta(\mathbf{r}) \ll 1$. 
\begin{align}
\delta{\bf A}_{\rm eff}(\alpha;x) &\equiv \delta{\bf p}(\alpha;x)=p_F \cos\alpha ~\delta \unitvec{l} \nonumber \\
&= -p_F \cos\alpha\,(\delta \unitvec{m} \cdot \unitvec{x})\hat{\bf z} = p_F \cos \alpha (\unitvec{y} \times \delta \unitvec{m}).
\label{A}
\end{align}
where $\unitvec{y} = \unitvec{z} \times \unitvec{x}$. 

Let us first consider the 2+1d fermions at $\alpha=0$. This is geometrically the 2+1d image of the Weyl point of $^3$He-A and $\mathbf{A}_{\rm eff}(0) = p_F\unitvec{l}_0$.
The only nonzero component of the 2+1d pseudomagnetic field is $B_y =\epsilon_{yzx} F_{zx}$ (i.e. the 2+1d magnetic field is along the neglected third dimension corresponding to the tangent $\unitvec{e}_{2}(0)=\unitvec{y}$),
\begin{align}
F_{ik}(\alpha= 0) = \frac{dA_i}{dx_k} - \frac{dA_k}{dx_i} =  \frac{dA_z}{dx}\hat{\bf z}_i \hat{\bf x}_k   \label{F}  \\
 \frac{dA_z}{dx} = p_F  \frac{d(\delta \unitvec{m}\cdot \unitvec{x})}{dx}=  p_F(\nabla\cdot\hat{\bf m})\,.
\label{B}
\end{align}
here the last equation follows if
\begin{align}
\delta \unitvec{m}(\mathbf{r}) \cdot \unitvec{x}  = \theta(\mathbf{r}) \equiv \theta(x) 
\end{align}
i.e. the deformation $\theta(\mathbf{r})$ only depends on $\mathbf{x}$. The covariant value of this pseudomagnetic feld is:
\begin{equation}
B^2(\alpha= 0)\equiv F_{ik}(0)F_{mn}(0)g^{im}g^{kn}= c_{\perp}^2v_F^2p_F^2 (\nabla\cdot\hat{\bf m})^2\,.
\label{FFpi2}
\end{equation}
The generalization of this expression to $\alpha \neq 0$ is
\begin{equation}
B^2(\alpha)= c_{\perp}^2v_F^2p_F^2 (\nabla\cdot\hat{\bf m})^2 \cos^4\alpha\,.
\label{FF}
\end{equation}
Here one factor of $\cos^2\alpha$ comes from the vector potential $\delta{\bf A}(\alpha)$ in Eq.(\ref{A}), and another factor of $\cos\alpha$ comes from the projection of $F_{zx}$ to the transverse $(\tilde x, z)$-plane normal to the nodal  line in Eqs. (\ref{F},\ref{B}), i.e. one has $\tilde{x}= x \cos\alpha - y \sin\alpha$, where $\tilde x$ is the local coordinate of the $2+1$d fermions perpendicular to the nodal line.

\subsection{Effective action for the polar pseudo EM field}

For 2+1d Dirac fermions, the Euler-Heisenberg effective action $\Delta L_{\rm QED}^{\rm 2+1d}[B]$ after integrating out the fermions is different from Eq. \eqref{FFS1} but instead is of the form \cite{Redlich1984, KatsnelsonVolovikZubkov2013}
\begin{align}
\Delta L^{\rm 2+1d}_{\rm grad}[B] = \sqrt{g_{\perp}} \frac{\zeta(3/2)}{4 \sqrt{2}\pi^2} \abs{B}^{3/2}. \label{Anomaly}
\end{align}
We note that in addition to this, a chiral 2+1d fermion may produce the parity/time-reversal anomaly related to the topological Chern-Simons term\cite{Redlich1984}. This is forbidden in the time-reversal invariant polar phase. In the chiral A-phase, in addition to \eqref{FFS1} one produces the topological Wess-Zumino term \cite{Volovik92} related to the chiral anomaly \cite{Bevan1997}, the coefficient of which changes sign after transition through the polar phase.

Integration of Eq. \eqref{FF} along the nodal line in momentum space, $p_F\int_0^{2\pi}  d\alpha~\abs{B^2(\alpha)}^{3/4}$, gives the non-analytic contribution to the gradient energy density
\begin{align}
\Delta F^{\rm polar}_{\rm grad}(B)= Cp_F \sqrt{-g_{\perp}}|B|^{3/2} \\
=C (c_{\perp}v_F)^{1/2}  (p_F)^{5/2}  |\nabla\cdot\hat{\bf m}|^{3/2} 
\,.
\label{F(B)}
\end{align}
where $g_{\perp} = 1/(v_F c_{\perp})$ is the determinant of the 2+1d metric in the dimensions perpendicular to the nodal line and
\begin{equation}
 C= \frac{\zeta(3/2)}{4\sqrt{2}\pi^2} \int_0^{2\pi} d\alpha~ \abs{\cos \alpha}^3  = \frac{\sqrt{2} \zeta(3/2)}{3\pi^2}\,.
\label{C}
\end{equation}
Substituting the physical parameters of the polar phase (the gap amplitude  $\Delta_P=c_{\perp}p_F$ and coherence length $\xi = v_F/\Delta_P$), one obtains
\begin{equation}
\Delta F^{\rm polar}_{\rm grad}=  \frac{\zeta(3/2)\sqrt{2}}{3\pi^2}v_F  p_F^2 \xi^{-1/2}  |\nabla\cdot\hat{\bf m}|^{3/2}    \,.
\label{F(B)2}
\end{equation}
The above expression resolves the singularity in the distorted A-phase in Eq.(\ref{FFSlimit}), when ${\bf e}_2\rightarrow 0$. The cross-over contribution between 3+1d and 2+1d in the polar distorted A-phase close to the transition to the polar phase can be approximated as (neglecting the logarithmic factor and leaving only the relevant terms):
\begin{equation}
\Delta F_{\rm grad} \sim   
 \frac{p_F^2v_F   }{ \frac{\xi^{1/2}} {(\nabla\cdot\hat{\bf m})^{3/2}} +
 \frac{  |{\bf e}_2|/|{\bf e}_1|}{( \hat{\bf l}\cdot (\hat{\bf l}\cdot \nabla)\hat{\bf m})^2}
 }
     \,.
\label{crossover}
\end{equation}

\subsection{Creation of quasiparticles by $2+1$d pseudoelectric field}

Adding a pseudoelectric field ${\bf E}= \partial_t{\bf A}_{\rm eff} = p_F \cos \alpha~ \partial_t(\unitvec{y}\times \delta \unitvec{m})$ leads to the substitution $\sqrt{B^2}\to \sqrt{B^2-E^2}$ in the above formulas \cite{Redlich1984, Volovik92}, as a result 
one obtains
\begin{align}
L_{\rm grad}[B,E] 
=   \sqrt{-g_{\perp}}\frac{\zeta(3/2)}{4\sqrt{2}\pi^2}v_F  p_F^2 \xi^{-1/2} \times \label{S} \\
\int_0^{2\pi} d\alpha \left((\nabla\cdot\hat{\bf m})^2 \cos^4\alpha- \frac{1}{v_F^2} (\partial_t \hat{\bf m})^2 \cos^2\alpha\right)^{3/4} \,. \nonumber
\end{align}
When pseudoelectric field is dominating, i.e. $B^2-E^2<0$, the expression in the brackets becomes negative and the action acquire a non-zero imaginary part, which describes the Schwinger effect, i.e. the production of pairs of  Dirac particles from the vacuum. For $B=0$, the probability of spontaneous creation of Dirac particles from the vacuum by the pseudoelectric field per unit time per unit volume  is:
 \begin{align}
{\rm Im}~ L_{\rm grad} =  {\sqrt{-g_{\perp}}}\frac{\zeta(3/2)}{4\sqrt{2}\pi^2} \sin(3\pi/4) \hspace*{.3\linewidth} \nonumber \\  
\times \frac{p_F^2}{v_F} \Delta_P^{-1/2}  |\partial_t \hat{\bf m}|^{3/2}
\int_0^{2\pi} d\alpha \abs{\cos\alpha}^{3/2}  \nonumber\\
=  \sqrt{-g_{\perp}} \frac{\zeta(3/2)}{4\pi^2} B\left(\frac{5}{4},\frac{1}{2}\right) \frac{p_F^2}{v_F} \Delta_P^{1/2}  |\partial_t \hat{\bf m}|^{3/2} .
\label{ImS}
\end{align}
where $B(x,y)$ is the beta function. The pseudoelectric field can be generated by the time dependent order parameter, which may result in dissipation proportional to  $\omega^{3/2}$, see Refs. \onlinecite{Volovik92, SchopohlVolovik92} for similar considerations for the A- and B-phases of $^3$He.

\section{Band touching line in a type-II Weyl material}

Finally, let us briefly discuss a topologically protected band touching or degeneracy surface of Weyl fermions related to the zero of the effective metric determinant. In the general case of interacting (or driven) systems, the Weyl fermions in the vicinity of the band touching are described by a 16 component of the tetrad field $e^{\mu}_{a}$, which enter the linear expansion of the $2\times 2$ Green's function \cite{NissinenVolovik2017b}, with $p_{\mu} = (\omega, \mathbf{p})$,
\begin{equation}
G^{-1}(p_\mu)=    e^0_0 \sigma^0 p_0+ e_0^i \sigma^0 p_i  +  e_a^i \sigma^a p_i +  e_a^0 \sigma^a p_0\,.
\label{GreensWeyGeneral}
\end{equation}
The tetrad components $e^i_0$ and $e^0_a$ are perturbations that parametrize the tilting of the Weyl cone in momentum and frequency space, leading to different types of Weyl 
fermions.\cite{Soluyanov2015,NissinenVolovik2017b} In the case of an isotropic Weyl cone characterized by a single ``speed of light" or Fermi-velocity 
$c$, the tetrads can be parametrized in terms of two velocity fields ${\bf v}$ and ${\bf w}$:
\begin{equation}
 e^0_0= -1
   \,\, , \,\, \, 
  e^i_a= c \delta^i_a
     \,\, , \,\, \, 
  e^i_0= v^i     \,\, , \,\, \, \,
       e^0_a=\frac{w_a}{c}  \,,
\label{SimplestTetrad}
\end{equation}
with the corresponding effective metric $g^{\mu\nu} = e^{\mu}_a e^{\nu}_{b}\eta^{ab}$,
\begin{equation}
  g^{\mu\nu}(\mathbf{v},\mathbf{w}) = \left(\begin{matrix} 1-\frac{w^2}{c^2} &  -(\mathbf{v}+\mathbf{w}) \\ -(\mathbf{v}+\mathbf{w}) & -c^2 \delta^{ij} + v^i v^j\end{matrix}\right) \,,
\label{SimplestMetric}
\end{equation}
and the determinant
\begin{align}
 g(\mathbf{v},\mathbf{w}) \equiv \det g_{\mu\nu} =-\frac{1}{e^2} = -\frac{1}{c^2 (c^2+{\bf v}\cdot{\bf w})^2} \\  
 e\equiv\det e_\alpha^\mu=-c(c^2+{\bf v}\cdot{\bf w})  \,.
\label{SimplestDeterminant}
\end{align}
The sign of the determinant of the tetrad field 
crosses zero either at $c=0$ or at $c^2 +{\bf v}\cdot{\bf w}=0$.
 The first case is similar to the crossing the polar phase, where the determinant changes sign when one of the speeds of light, the speed of light propagating   along ${\bf e}_2$, crosses zero. Let us consider the second case.

The energy-momentum dispersion $g^{\mu\nu}p_{\mu}p_{\nu}=0$ is given by the poles of the Green's function, $G^{-1}(\omega,\mathbf{p}) =0$ and it consists of the two bands:
\begin{widetext}
\begin{align}
\omega_{\pm}(\mathbf{p}) = \frac{\mathbf{p}\cdot (\mathbf{v}+\mathbf{w}) \pm \sqrt{(1-\frac{w^2}{c^2})(c^2 p^2 -(\mathbf{p}\cdot \mathbf{v})^2)+(\mathbf{p}\cdot(\mathbf{v}+\mathbf{w}))^2}}{1-\frac{w^2}{c^2}} \label{eq:poles}
\end{align}
\end{widetext}
The two bands are degenerate when the square root in Eq.(\ref{eq:poles}) is zero, i.e. at
\begin{equation}
\left( w({\bf p}\cdot {\bf v}) + c^2p_\parallel\right)^2+ (c^2-w^2)c^2{\bf p}_\perp^2=0 \,,
\label{TouchningCondition}
\end{equation}
where $p_\parallel$ and ${\bf p}_\perp$ are components of momentum parallel and transverse to ${\bf w}$ respectively. Let us consider the case when $w<c$ and $v>c$, which corresponds to an overtilted or type-II Weyl fermions with a frequency-momentum normalization $\mathbf{w}$. Then Eq.(\ref{TouchningCondition}) is satisfied when ${\bf p}_\perp=0$ and $c^2 +{\bf v}\cdot{\bf w}=0$, i.e. at the transition, at which the determinant of the tetrad changes sign and the right-handed fermions transform to the left-handed ones \cite{Rovelli2012a,Rovelli2012b}. At this transition the two bands touch each other at the line 
${\bf p}\times {\bf w}=0$, where $\omega_+=\omega_- = - c^2({\bf p}\cdot {\bf w}) /w^2$.  In contrast to the Dirac nodal line in the polar phase of $^3$He, here the band touching or degeneracy does not occur at zero-energy (i.e. at the chemical potential). For the latter to occur, a symmetry protecting the degeneracy at zero energy is required, such as chiral symmetry and/or time reversal symmetry in the polar phase of $^3$He. Nevertheless, the surface of band touching is still topologically protected, see for example \cite{Heikkila2015,Heikkila2016}.

\section{Conclusion }

In this paper, we considered the effective 3+1d and 2+1d QED coupled to massless Weyl and Dirac fermions and the associated Euler-Heisenberg effective actions in the orbital dynamics of polar distorted superfluid $^3$He-A. In particular, one can continuously dial between the 3+1d Weyl and 2+1d Dirac fermions by tuning the orbital structure of the order parameter \cite{Dmit1}.  
 
We considered the transition in terms of the effective spacetime experienced by the low-energy fermions via the emergent tetrad fields. In this condensed matter setting, one can realize the transition from "spacetime to anti-spacetime" \cite{Rovelli2012b}. In superfluid $^3$He the change of the determinant of the Weyl fermion tetrad field occurs through the topological Lifshitz transition of the Bogoliubov-Fermi line associated with the polar phase with a degenerate zero tetrad component.
While the action for fermions is different in spacetime and anti-spacetime, the effective action for the bosonic fields (effective electromagnetic and gravity fields) is the same for both tetrads. The bosonic action is proportional to $\sqrt{-g}=|\det e|$ and this is a non-analytic function of the determinant of the tetrad field $\det e$. The nonanalytical behavior is demonstrated at the transition, where the tetrad field becomes degenerate. This leads to the topologically protected Dirac nodal line in the polar phase and to the band touching surface in Weyl semimetals. When the Lifshitz transition is approached the bosonic action develops a singularity, at which the $3+1$d action effectively transforms to a one-parameter family of $2+1$d Dirac fermions.
In particular, the Schwinger pair production of massless fermions by the effective electric field, which is proportional  to $E^2$ in $3+1$d quantum electrodynamics and in Weyl superfluid $^3$He-A, transforms  to the $E^{3/2}$ behavior in the polar phase, which will be a definitive characteristic of the $2+1$ quantum electrodynamics emerging near the nodal Dirac line.

We note that we focused on the Euler-Heisenberg effective actions and did not consider any possible topological terms of the effective action that are related to anomalies: the chiral anomaly and Wess-Zumino term for $^3$He-A and the parity anomaly for the 1d family of 2+1d Dirac 
fermions.\cite{RuiZhaoSchnyder17} The latter is not relevant for the 3+1d polar phase which is time-reversal invariant, while the Wess-Zumino term in distorted $^3$He-A changes sign after transition through the polar phase.

The original fermionic action is analytic and remains analytic across the Lifshitz transition through the superfluid states corresponding to the degenerate tetrad. The fermionic action can be rewritten in the fully covariant way, but after that it becomes nonanalytic in the tetrad field. This nonanalyticity is extended to the effective bosonic action and
is a natural consequence of the emergent quasirelativistic symmetry of the low-energy condensed matter system. This gives one suggestive answer to the question posed by Diakonov\cite{Diakonov2011} and Rovelli {\it et al.}\cite{Rovelli2012b}:
``Anti-spacetimes" with $\det\, e<0$ may exist and they contribute trivially to quantum gravity in a sense that the emergent bosonic action does not resolve between spacetime and anti-spacetime, since it is proportional to $|\det e|$. This is in contrast to topological terms in the effective action which arise often due to anomalous discrete transformations relating spacetime and anti-spacetime. 

This work has been supported by the European Research Council (ERC) under the European Union's Horizon 2020 research and innovation programme (Grant Agreement No. 694248). GV thanks M. Zubkov for numerous discussions.

\end{document}